# Parametric amplification and self-oscillation in a nanotube mechanical resonator

*Alexander Eichler, Julien Chaste, Joel Moser, Adrian Bachtold[*]*

Catalan Institute of Nanotechnology (ICN) and CIN2, Campus UAB, 08193 Bellaterra, Barcelona, Spain

* adrian.bachtold@cin2.es

A hallmark of mechanical resonators made from a single nanotube is that the resonance frequency can be widely tuned. Here, we take advantage of this property to realize parametric amplification and self-oscillation. The gain of the parametric amplification can be as high as 18.2 dB and tends to saturate at high parametric pumping due to nonlinear damping. These measurements allow us to determine the coefficient of the linear damping force. The corresponding damping rate is lower than the one obtained from the lineshape of the resonance (without pumping), supporting the recently reported scenario that describes damping in nanotube resonators by a nonlinear force. The possibility to combine nanotube resonant mechanics and parametric amplification holds promise for future ultra-low force sensing experiments.

Carbon nanotubes allow the fabrication of nanoelectromechanical resonators with outstanding properties. The resonance frequency can be very high [1] and at the same time widely tunable [2-3]. In addition, nanotube resonators are very sensitive to electron charges [4,5], to mass [6-8] and to force [9]. A major issue in these experiments is the detection of the motion. The difficulty lies in transducing the high-frequency subnanometer amplitude of the motion into a sizeable electrical signal. This is especially true for ultra-sensitive force sensing experiments where the transduction has to be as efficient as possible. Usually, the mechanical motion is directly converted into a voltage which subsequently undergoes amplification with high gain. A strategy to improve the detection sensitivity is to preamplify the motional amplitude before the electrical conversion using the parametric effect [10].

Parametric amplification in mechanical resonators has been intensively studied [10-22]. It is not only employed to amplify mechanical signals, it also allows for the enhancement of the quality factor [13,17,22], the storage and the operation of mechanical bits [16], thermal noise squeezing [10,20-21], and the reduction of the parasitic signal in capacitive detection schemes [11]. In its most conventional form, parametric amplification consists of modulating the resonator spring constant $k_0$ at twice the resonant frequency $f_0$ [23]. This is achieved in many NEMS resonators by tuning $k_0$ electrostatically with a voltage $V_g$ applied on a nearby gate. Nanotube resonators are expected to be excellent candidates for parametric amplification because $k_0$ can be modulated with $V_g$ by a very large amount: the modulation can be made larger than in any other mechanical resonators fabricated to date (this can be quantified by measuring the $V_g$ dependence of the resonance frequency, which scales as $\sqrt{k_0}$). However, parametric amplification in a nanotube resonator could not be realized thus far. One reason for this is that the employed transduction schemes [3,24] are not suitable for such measurements (see below).

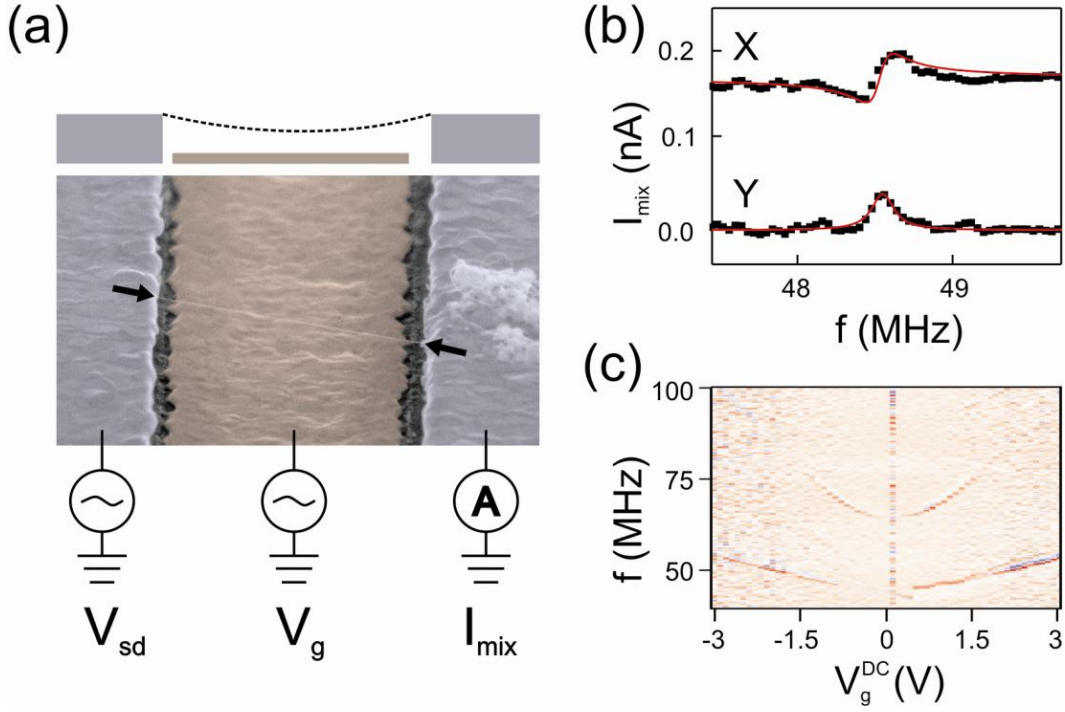

**Figure 1: (a)** Schematic diagram and false-colour scanning electron microscopy (SEM) image of the device. The nanotube (arrows, dashed line in the SEM image) is suspended over a gate electrode (red) between two metal electrodes (gray). The distance between the electrodes is 1 μm. All measurements are performed in an ultra-high vacuum chamber (about $10^{-10}$ mbar) at 100 K (in order to avoid the interplay between parametric amplification and Coulomb blockade which emerges at lower temperature). **(b)** In-phase component $X$ and out-of-phase component $Y$ of $I_{mix}$ (without parametric pumping). The motional amplitude at the resonance frequency is proportional to $Y$. Red lines are fits using the real and imaginary parts of a Lorentzian. $V_g^{AC} = 5$ mV, $V_{sd}^{AC} = 1.4$ mV. We estimate the motion amplitude to be ~ 2 nm by comparing the on-resonance signal of $Y$ with the off-resonance signal of $X$ in conjunction with eq. 2 as in [3]. In comparison, the amplitude of the thermal motion is $\sqrt{k_B T/m(2\pi f_0)^2} = 1.4$ nm ($k_B$ is the Boltzmann constant and we take $m = 7 \cdot 10^{-21}$ kg the mass of the suspended nanotube). **(c)** Resonance frequency as a function of gate voltage (from measuring $I_{mix}$ versus $f$ and $V_g^{DC}$). Two mechanical modes can be seen. All results that follow are obtained for $V_g^{DC} = -1.5$ V and $f \sim 50$ MHz.

Here, we report on a new version of the mixing technique [3] that detects the two quadratures of the motion of a nanotube resonator. This allows us to study the parametric amplification of the motion and to demonstrate a gain as high as 18.2 dB. When $k_0$ is modulated above a threshold value, the nanotube is shown to enter a regime of instability and self-oscillation; that is, the nanotube oscillates even though the driving force is set to zero.

Our nanoresonators consist of a suspended carbon nanotube clamped between two metal electrodes (Fig. 1a) and are fabricated as follows: a trench is etched into a highly resistive Si wafer coated with $SiO_2$ and $Si_3N_4$, and W and Pt are evaporated into the trench to create a gate electrode. In a second lithography step, a continuous line is exposed across the trench. After deposition of W/Pt and lift-off, the line results in the source and drain electrodes separated by the trench (these electrodes are electrically isolated from the gate due to the undercut profile of the $Si_3N_4/SiO_2$ substrate). An island of catalyst is patterned on the drain (or source) electrode using electron-beam lithography and nanotubes are grown by chemical vapor deposition. This growth is the last step of the fabrication process, and therefore the nanotubes are not contaminated with residues of the resists and chemicals [25-27].

The nanoresonator is actuated by applying a voltage $V_g^{AC}$ at frequency $f$ to the gate, which causes a driving force $F \propto V_g^{AC}$. The resulting motion is detected by applying a voltage $V_{sd}^{AC}$ at a slightly detuned frequency ($f - \delta f$) to the source electrode, and by measuring the mixing current $I_{mix}$ at frequency $\delta f$ from the drain electrode using a lock-in amplifier. In previous works [3,4,6,7], the recorded signal was the modulus of $I_{mix}$, which reads

$$I_{mix} = \frac{1}{2} V_{sd}^{AC} \frac{\partial G}{\partial V_g} \left( V_g^{AC} \cos(2\pi\delta f t - \varphi_E) + z_0 V_g^{DC} \frac{C'_g}{C_g} \cos(2\pi\delta f t - \varphi_E - \varphi_M) \right) \quad (1)$$

Here, $z_0$ is the motional amplitude, $\frac{\partial G}{\partial V_g}$ the transconductance, $t$ the time, $V_g^{DC}$ the constant voltage applied to the gate, $C_g$ the gate-nanotube capacitance, and $C'_g$ its derivative with respect to the displacement. The phase $\varphi_M$ is the phase difference between the displacement and the driving force, and $\varphi_E$ is the phase difference between the $V_{sd}^{AC}$ and $V_g^{AC}$ signals. In practice, $\varphi_E$ is difficult to control (it depends on the details of the measurement circuit such as the cable lengths). Because two phases are at work, the measurement of the modulus of $I_{mix}$ is not convenient to extract $z_0$. Moreover, we find that $\varphi_E$ can change with the power of the applied oscillating voltages, which is not suitable for the study of parametric amplification. Other versions of the mixing technique are not appropriate for such a study either. In the frequency-modulation technique [24], $I_{mix}$ is proportional to the derivative of $z_0$ with respect to $f$. In the amplitude-modulation technique [28], $I_{mix}$ does not measure $z_0$ at $f_0$ (since $I_{mix}$ depends only on $\mathrm{Re}[\tilde{z}(f)]$).

We revisit the mixing technique to measure the two quadratures of the motion, $\mathrm{Re}[\tilde{z}(f)]$ and $\mathrm{Im}[\tilde{z}(f)]$. When actuating the resonator with an oscillating force at frequency $f$, the displacement can be written as $z = \mathrm{Re}[\tilde{z}(f)]\cos(2\pi f t) + \mathrm{Im}[\tilde{z}(f)]\sin(2\pi f t)$ and $I_{mix}$ can take the form (see supplementary information)

$$I_{mix} = \frac{1}{2} V_{sd}^{AC} \frac{\partial G}{\partial V_g} \left( V_g^{AC} \cos(2\pi\delta f t - \varphi_E) + V_g^{DC} \frac{C'_g}{C_g} \mathrm{Re}[\tilde{z}(f)]\cos(2\pi\delta f t - \varphi_E) + V_g^{DC} \frac{C'_g}{C_g} \mathrm{Im}[\tilde{z}(f)]\sin(2\pi\delta f t - \varphi_E) \right). \quad (2)$$

For a properly tuned phase of the lock-in amplifier, the out-of-phase component of the lock-in amplifier output, $Y$, corresponds to the imaginary part of the resonance (third term in eq 2) and the in-phase component, $X$, to the real part of the resonance (second term in eq 2) added to a constant background which has a purely electrical origin (first term in eq 2). In order to find this phase, we choose a driving frequency far from $f_0$ and tune the phase of the lock-in amplifier until the measured out-of-phase component is zero. Figure 1b shows the two quadratures of $I_{mix}$ when sweeping $f$. The red line is a fit with eq 2 of the measurements assuming that the resonator is described as a damped harmonic oscillator. At the resonance frequency, $Y$ is directly proportional to the amplitude of motion ($z_0 = \text{Im}[\tilde{z}(f_0)]$ and $\text{Re}[\tilde{z}(f_0)] = 0$), which is very practical to study parametric amplification.

Before discussing parametric amplification, we characterize the $V_g^{DC}$ dependence of $k_0$. For this, we measure $f_0$ as a function of $V_g^{DC}$ (since $k_0 \propto f_0^2$). Figure 1c shows two clearly resolved resonant modes. Their resonance frequency can be tuned to a large extent with $V_g^{DC}$. This behavior has been attributed to the increase of the elastic tension that builds up in the nanotube as it bends towards the gate with increasing $V_g^{DC}$ [3]. For the lower frequency mode, $df_0/dV_g$ is constant to a good accuracy over several volts and equal to 4.9 MHz/V. Compared to other NEMS resonators, this response is exceptionally high. Previous studies attained up to $df_0/dV_g = 2$ kHz/V with capacitive forces [10,13], 240 kHz/V using the Lorentz force [19], up to 40 kHz/V for piezoelectric NEMSs [16-17,21-22], and 10 kHz/V using a dielectric force setup [18]. Recently, $df_0/dV_g = 1.1$ MHz/V was obtained by coupling a resonator to a Cooper pair box at very low temperature (130 mK) [20]. The high values of $df_0/dV_g$ achieved in nanotube resonators (up to 10 MHz/V at room temperature in [3]) makes them excellent candidates for parametric amplification.

In order to realize parametric amplification, we apply an additional oscillating voltage $V_P$ at a frequency $2f$ to the gate. On resonance, this modulates $k_0$ at $2f_0$, thereby achieving parametric pumping of the resonator. We measure the resulting amplification of $z_0$ by comparing $Y_{pumped}$ to the unpumped signal $Y_{unpumped}$. Figure 2a shows that the mechanical amplification $\Lambda = Y_{pumped}/Y_{unpumped}$ depends on the phase $\Delta\phi$ of the driving force with respect to the pump excitation. The maximum amplification is achieved at about $\Delta\phi = -45°$. We plot the gain corresponding to the amplification at this phase as a function of $V_P$ in Fig. 2b. The largest gain is 18.2 dB and is obtained for $V_P = 11.5$ mV. Beyond this pumping voltage, we find that the signal becomes highly unstable.

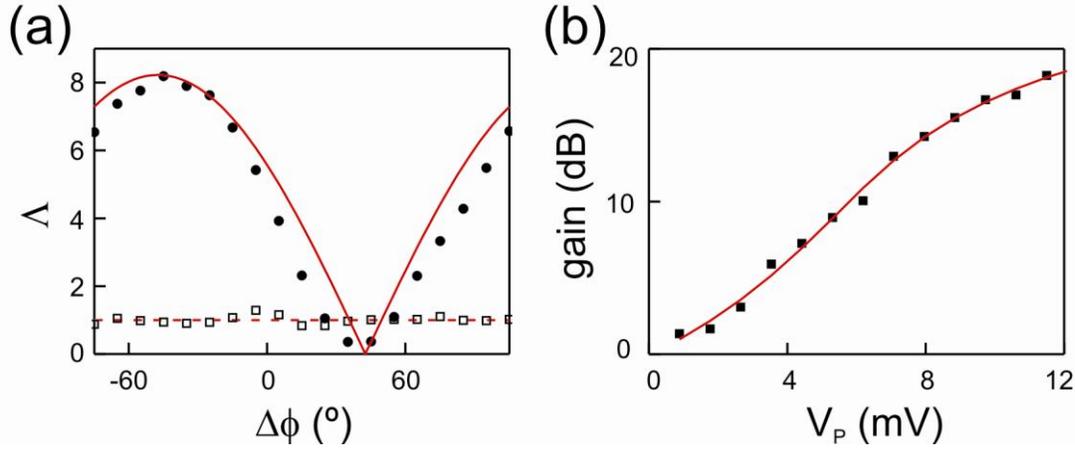

**Figure 2: (a)** Amplification ($\Lambda = Y_{pumped}/Y_{unpumped}$) as a function of the phase difference $\Delta\phi$ between the driving force and the pump for $V_P = 11.5$ mV (solid dots) and $V_P = 0$ (hollow squares). $V_g^{AC} = 3$ mV and $V_{sd}^{AC} = 1.4$ mV. The solid red line is a fit with eq 3. The dashed red line is the mean value when the pump is off. Our measurement setup can only control the shift of $\Delta\phi$; we have determined $\Delta\phi = 0$ from optimizing the agreement between experiment and theory. **(b)** Gain as a function of the amplitude of $V_P$ for $\Delta\phi$ at which the amplification is largest. The solid red line is a fit with eq 4. $V_g^{AC} = 3$ mV and $V_{sd}^{AC} = 1.4$ mV. We estimate the motion amplitude to be ~ 10 nm at the largest gain.

We analyze our data in the framework of the parametric excitation of a damped resonator. From reference [23], we find that the amplification is

$$\Lambda = \left| \text{Im}\left[ -e^{i\pi/4}\left( \frac{\cos(\Delta\phi + \pi/4)}{1 - V_P/V_{P,C}} + i\frac{\sin(\Delta\phi + \pi/4)}{1 + V_P/V_{P,C}} \right) \right] \right|. \tag{3}$$

Here, $V_{P,C}$ is the critical pumping voltage for which the amplification is expected to diverge ($V_{P,C} = (f_0 \cdot dV_g/df_0)/Q_0$ with $Q_0$ the quality factor associated to the damping force $\gamma \dot{z}$, where $\dot{z}$ is the velocity). We compare the measurements to eq 3 using $V_{P,C}$ as a fitting parameter (Fig. 2a, solid red line). The agreement is reasonable and we obtain $V_{P,C} = 12.5$ mV. (We note that the linear model used for eq. 3 is not strictly valid for large $V_P$, see below.)

Regarding the $V_P$ dependence of the amplification in Fig. 2b, the measured amplification tends to saturate at high $V_P$. This is in opposition to the divergent growth expected from eq 3, which assumes that damping is described by the linear force $\gamma \dot{z}$. According to the theory of parametric amplification [23], the saturation can be accounted for by adding a nonlinear damping force $\eta z^2 \dot{z}$ to the Newton equation of a harmonic oscillator. This force leads to saturation because the associated energy dissipation depends on the amplitude, i.e. higher amplitudes correspond to higher loss of energy. This nonlinear damping force, which naturally emerges from a nonlinear Caldeira-Leggett model [29], has recently been shown to be crucial to explain the measured driven resonances of nanotube resonators [9] (more discussion on this force can be found below). Following previous work [20, 23] the saturation can be quantified by finding the solution $\Lambda$ of the equation

$$V_P = u\Lambda^2 - \frac{v}{\Lambda} + V_{P,C}. \tag{4}$$

Here, $u = \frac{\pi \eta Q_0 f_0 V_{P,C}}{k_0} \text{Im}[\tilde{z}(\omega)]^2_{unpumped}$ and $v = \frac{Q_0 F V_{P,C}}{\sqrt{2} k_0 \text{Im}[\tilde{z}(\omega)]_{unpumped}}$, where $F$ denotes the driving force. The measurements are compared to the solution of eq 4 using $u$, $v$, and $V_{P,C}$ as fitting parameters (solid red line in Fig. 2b). The agreement is good and we obtain $V_{P,C} = 7.5$ mV, which is rather similar to the value found above.

Upon increasing the pump excitation, the nanotube is observed to self-oscillate (Fig. 3). That is, the nanotube resonator enters in a regime where it vibrates without any driving force: an immobile resonator is expected to be instable when parametrically pumped with a voltage above $V_{P,C}$, and any fluctuation will activate the oscillating motion [23]. In the measurements, we set $F = 0$ (by putting the oscillating voltage $V_g^{AC}$ at frequency $f$ to zero) and we measure the mixing current as a function of $V_P$ and pump frequency. Figure 3a shows mechanical motion in a tongue-shaped region, which is a typical signature of self-oscillation [11,15-16,20,22-23]. When sweeping the pump frequency in the opposite direction, the measurement is different (Fig. 3b); the hysteresis is attributed to a (negative) Duffing force [16]. In both sweep directions, self-oscillation is observed for $V_P$ roughly above $V_{P,C} = 10$ mV. On a technical note, the motion is detected because we apply the voltage $V_{sd}^{AC}$ at frequency ($f - \delta f$) with $\delta f = 10$ kHz. Although $\delta f$ is lower than the resonance width (about 150 kHz at low drive, see below), it is unlikely that the voltage $V_{sd}^{AC}$ with frequency $f - \delta f$ affects the measurements in Fig. 3 for the following reasons: firstly, $V_{sd}^{AC}$ is low (1.4 mV) and the corresponding force is not enough to actuate the resonator in a detectable way (the resonance is detected when the drive is equal to or larger than 3 mV). Secondly, the nanotube resonator behaves as expected in the self-oscillation regime. Namely, we observe no mixing current when the pumping voltage is below the threshold value $V_{P,C}$ and, in addition, the region in which we detect motion has the tongue-shape characteristic of parametric self-oscillation.

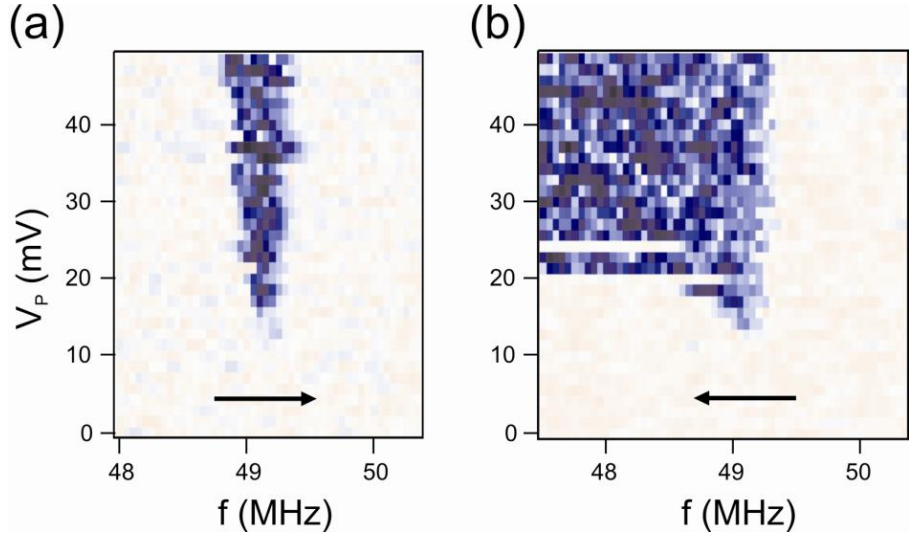

**Figure 3:** Parametric self-oscillation (blue) when the pumping frequency is swept upwards **(a)** and downwards **(b)** (by measuring $I_{mix}$ as a function of the detection frequency $f$ and $V_P$). The driving force at frequency $f$ is set to zero. The observed hysteresis is due to a (negative) Duffing force. In these frequency sweeps, the phase difference between $V_P$ and $V_{sd}^{AC}$ is not kept fixed so the measured $I_{mix}$ is fluctuating. $V_{sd}^{AC} = 1.4$ mV.

In these measurements of parametric amplification and self-oscillation, we obtain three individual estimations of $V_{P,C}$ that are rather similar (about 10 mV). Using $Q_0 = (f_0 \cdot \partial V_g / \partial f_0)/V_{P,C}$ this corresponds to a quality factor of about 1000. Surprisingly, this is significantly larger than the quality factor obtained when the pump is off (e.g. in Fig. 1b where the quality factor is extracted by comparing the resonance lineshape with the predictions of a damped harmonic oscillator). There, the quality factor is about 170-350 (crosses in Fig. 4a). Measurements on a second resonator give the same results (see supporting information). We attribute this difference to the different damping forces that are at work: $\gamma \dot{z}$ and $\eta z^2 \dot{z}$. The experiments on parametric amplification and self-oscillation are sensitive to the critical pump excitation $V_{P,C}$, which is a direct measure of $\gamma$. However, $V_{P,C}$ does not necessarily quantify the total damping in the resonator. Indeed, we recently showed [9] that the principal contribution to the damping in a nanotube resonator can stem from the $\eta z^2 \dot{z}$ force. In this case, the quality factor $Q$ estimated from the resonance lineshape is lower than $Q_0 = k_0/2\pi f_0 \gamma$ obtained from $V_{P,C}$, which is in agreement with our findings. In the following, we give further experimental evidences that the damping in the studied resonator indeed emanates from the $\eta z^2 \dot{z}$ force. Figures 4b,c show that the measured resonance lineshape compares reasonably well with the predictions of a Duffing resonator with nonlinear damping, the equation of motion being described by [9]

$$m\ddot{z} = -k_0 z - \gamma \dot{z} - \alpha z^3 - \eta z^2 \dot{z} + F\cos(2\pi f t). \qquad (5)$$

The whole set of resonance lineshapes measured at different driving forces can be fitted with a single value for $\alpha$ and $\eta$. The corresponding value of $Q$ depends on the driving force (squares in Fig. 4a), which signals that the damping is nonlinear, as demonstrated in [9].

An important step forward would be to optimize nanotube resonators in order to further enhance the parametric gain. Since the theory of parametric amplification predicts that the gain is limited by nonlinear damping [23], methods to reduce this nonlinear damping force should be developed. This is of primary importance since the same nonlinear damping also sets the quality factor of driven nanotube resonators [9]. The microscopic origin of the nonlinear damping is not clear, but it could be related to phonon tunneling, sliding at the contacts, nonlinearities in phonon-phonon interactions, or contamination in combination with geometrical nonlinearities [9]. We will experimentally study the dependence of the nonlinear damping force on contamination, the clamping configuration, and the suspended length. Theoretical work on the microscopic nature of nonlinear damping will prove useful [29,30].

In conclusion, we demonstrate parametric amplification and self-oscillation in a carbon nanotube resonator. Our results hold promise for ultra-low force sensing experiments. We recently demonstrated a force sensitivity of 2.5 aN·Hz$^{-1/2}$ with a nanotube resonator (without parametric amplification) [9] and if it were possible to achieve with this resonator the same gain as reported in the present work, this would surpass the record force sensitivity of 0.51 aN·Hz$^{-1/2}$ recently demonstrated in Ref. [31]. Moreover, nanotube mechanical resonators are promising systems for future studies of the interplay between parametric amplification and Coulomb blockade [32].

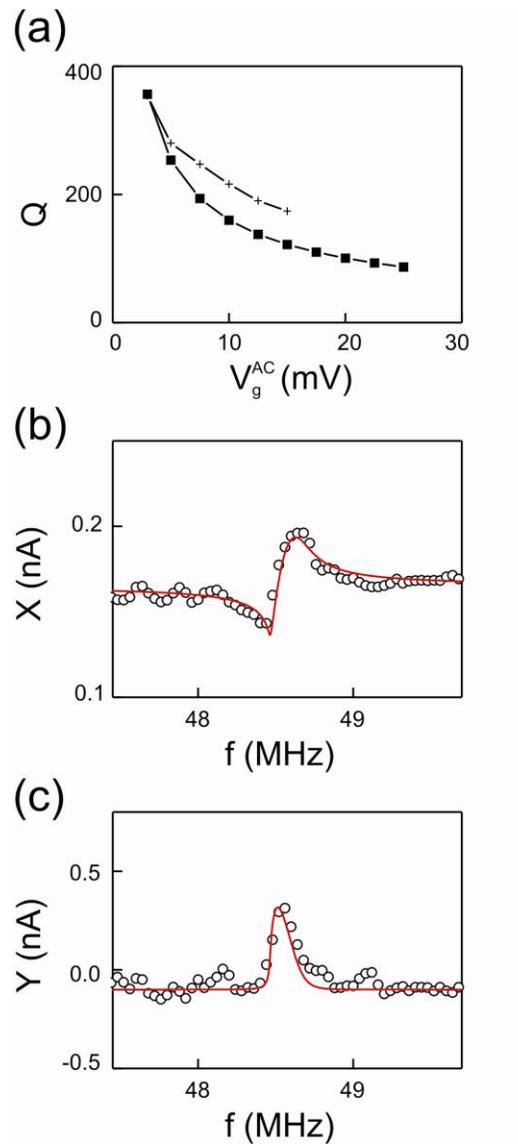

**Figure 4:** **(a)** Quality factor as a function of the driving voltage $V_g^{AC}$ in the absence of parametric

pumping. The crosses correspond to the quality factor obtained by fitting the resonance lineshape with the predictions of a damped harmonic oscillator. The filled squares are obtained by fitting the resonance lineshapes of $Y$ with the solution of eq 5 with a single value for $\alpha$ and $\eta$ for the different driving forces ($\eta = 120$ kg·m$^{-2}$s$^{-1}$, $\alpha = -1.4 \cdot 10^{10}$ kg·m$^{-2}$s$^{-1}$, $\gamma = 0$). **(b)** and **(c)** Comparison of the resonance lineshapes of the in-phase component $X$ and the out-of-phase component $Y$ of $I_{mix}$ with the solution of eq 5. $V_g^{AC} = 5$ mV and $V_{sd}^{AC} = 1.4$ mV.

ACKNOWLEDGMENT We thank I. Wilson-Rae for a critical reading of the manuscript. We acknowledge support from the European Union (RODIN, FP7), the Spanish ministry (FIS2009-11284), the Catalan government (AGAUR, SGR), the Swiss National Science Foundation (PBBSP2-130945), and a Marie Curie grant (271938). We thank Brian Thibeault (Santa Barbara) for help in fabrication.

SUPPORTING INFORMATION PARAGRAPH: A derivation of equations 2, 3, and 4 is provided. Measurements of self-oscillation from a second resonator are presented.

Supporting Information for

**Parametric amplification and self-oscillation in a nanotube mechanical resonator**

A. Eichler[1], J. Chaste[1], J. Moser[1], and A. Bachtold[1]

[1]Catalan Institute of Nanotechnology (ICN) and CIN2, Campus UAB, 08193 Bellaterra, Barcelona, Spain

Equations 3 and 4 in the main text are drawn from reference [1]. Please note that we use a different notation than in [1] in order to reserve certain symbols for physical quantities.

## A) Derivation of equation 2

We apply $V_g^{AC}(t) = V_g^{AC}\cos(\omega t)$ with $\omega = 2\pi f$. The resulting force $F\cos(\omega t)$ induces an oscillation of the nanotube position $z = \text{Re}[\tilde{z}(\omega)]\cos(\omega t) + \text{Im}[\tilde{z}(\omega)]\sin(\omega t)$ with

$$\tilde{z}(\omega) = \pi F / m(\omega_0^2 - \omega^2 - i\omega_0\omega/Q) + \pi F / m(\omega_0^2 - \omega^2 + i\omega_0\omega/Q) \tag{S0}$$

and an oscillation of the nanotube conductance [2]

$$\delta G = \frac{\partial G}{\partial V_g}\left(V_g^{AC}\cos(\omega t) + z V_g^{DC}\frac{C_g'}{C_g}\right). \tag{S1}$$

When applying $V_{sd}^{AC}(t) = V_{sd}^{AC}\cos((\omega - \delta\omega)t + \varphi_E)$, we get

$$I_{mix} = \frac{1}{2}V_{sd}^{AC}\frac{\partial G}{\partial V_g}\left(V_g^{AC}\cos(\delta\omega\cdot t - \varphi_E) + V_g^{DC}\frac{C_g'}{C_g}\cos(\delta\omega\cdot t - \varphi_E)\text{Re}[\tilde{z}(\omega)] + V_g^{DC}\frac{C_g'}{C_g}\sin(\delta\omega\cdot t - \varphi_E)\text{Im}[\tilde{z}(\omega)]\right)$$
(S2)

at frequency $\delta\omega$.

## B) Derivation of equation 3

As we explain in the main text, we tune the phase of the lock-in amplifier with which we measure the mixing current, such that $X \propto \text{Re}[\tilde{z}(\omega)]$ and $Y \propto \text{Im}[\tilde{z}(\omega)]$. The secular perturbation theory in [1] employs dimensionless variables that are related to the physical ones by

$$\xi = z\sqrt{\frac{\alpha}{m\omega_0^2}}; \quad G = \frac{F}{\omega_0^3}\sqrt{\frac{\alpha}{m^3}}; \quad \bar{t} = \omega_0 t; \quad \text{and} \quad \bar{\omega} = \frac{\omega}{\omega_0}; \tag{S3}$$

where $\alpha$ denotes the coefficient of the Duffing cubic force, $m$ the resonator mass, $F$ the coefficient of the driving force $F\cos(\omega t)$, and $\omega_0 = 2\pi f_0$. The other variables are defined in the main text.

In a next step, a complex amplitude $A(T)$ is introduced, where $T = \varepsilon \cdot \bar{t}$ is a slow time variable and $\varepsilon = 1/Q_0$ ($Q_0 = m\omega_0/\gamma$ is the quality factor, $\gamma$ being the linear damping constant). Following [1] we use the ansatz

$$\xi(\bar{t}) = \frac{\sqrt{\varepsilon}}{2}(A(T)\cdot e^{i\bar{t}} + c.c.), \tag{S4}$$

where $c.c.$ denotes complex conjugation. Assuming a steady-state solution of the form

$$A(T) = ae^{i\Omega T} = |a|e^{i\phi}e^{i\Omega T} \tag{S5}$$

this leads to the expressions

$$\xi(\bar{t}) = |a|\sqrt{\varepsilon}\cos(\bar{\omega}\cdot\bar{t} + \phi) \tag{S6}$$

$$z(t) = |a|\sqrt{\gamma\omega_0/\alpha}\cos(\omega t + \phi). \tag{S7}$$

Using $a = \text{Re}[a] + i\,\text{Im}[a]$ and $e^{i(\Omega T + \bar{t})} = \cos(\Omega T + \bar{t}) + i\sin(\Omega T + \bar{t})$, we get that

$$z(t) = \sqrt{\gamma\omega_0/\alpha}(\text{Re}[a]\cos(\omega t) - \text{Im}[a]\sin(\omega t)) \tag{S8}$$

Without pumping, we have at resonance (defined as the frequency for which the motional amplitude is largest) $\text{Re}[a] = 0$ and $|\text{Im}[a]| = |g|$ where $g = G\varepsilon^{-3/2}$ (using eq. (1.30) of [1] and assuming that the nonlinear damping force is negligible), so

$$X_{unpumped} = 0 \text{ and } Y_{unpumped} = r\cdot g \tag{S9}$$

with $r$ a real constant (using eq. S2 and S8).

When the pumping is on (i.e. the spring constant is modulated as $k(1+H\cos(\omega_p t))$), eq. (1.52) of [1] reads

$$a = -e^{i\pi/4}\left(\frac{\cos(\Delta\phi+\pi/4)}{1-h/2} + i\frac{\sin(\Delta\phi+\pi/4)}{1+h/2}\right)|g| \qquad (S10)$$

where $\Delta\phi$ is the phase of the driving force with respect to the pumping and $h/2 = V_P/V_{P,C}$ (here $h = H/\varepsilon = \frac{2Q_0}{f_0}\frac{df_0}{dV_g}V_P$ and $V_{P,C} = (f_0 \cdot dV_g/df_0)/Q_0$). Please note that the equation appears in [1] without a minus sign. We measure at resonance

$$Y = r \cdot \text{Im}[a] \ . \qquad (S11)$$

Using eq. S9, S10, and S11, we obtain

$$\left|\frac{Y_{pumped}}{Y_{unpumped}}\right| = \left|\text{Im}\left[-e^{i\pi/4}\left(\frac{\cos(\Delta\phi+\pi/4)}{1-V_P/V_{P,C}} + i\frac{\sin(\Delta\phi+\pi/4)}{1+V_P/V_{P,C}}\right)\right]\right| \qquad (S11)$$

**C) Derivation of equation 4**

Introducing the nonlinear damping force $\eta z^2 \dot{z}$ in the Newton equation, Lifshitz and Cross obtained (eq. 1.70 of [1])

$$\frac{db}{dT} = \frac{1}{2}\frac{h-h_C}{h_C}b - \frac{\sigma}{8}b^3 + \frac{|g|}{2}\cos(\Delta\phi+\pi/4) \qquad (S12)$$

where $b = Ae^{-i\pi/4}$ is a real constant, $\sigma = \frac{\eta\omega_0}{\alpha}$, and $h_C = \frac{2Q_0}{f_0}\frac{df_0}{dV_g}V_{P,C}$. Please note that the last term on the right-hand side has a different sign in [1] (because of the minus sign in eq. S10). Following [1], we are interested in a time-independent solution ($db/dT=0$) at maximum gain ($\Delta\phi = -\pi/4$). At resonance, we have $\Omega = 0$. We require a solution for $\text{Im}[a] = \text{Im}[A] = b/\sqrt{2}$, which satisfies

$$h = \frac{1}{2} h_C \sigma \operatorname{Im}[a]^2 - \frac{1}{\sqrt{2}} \frac{|g| h_C}{\operatorname{Im}[a]} + h_C. \tag{S13}$$

After inserting the physical units and using eq. S3 and S8, we get

$$V_P = \frac{\pi \eta Q_0 f_0 V_{P,C}}{k_0} \operatorname{Im}[\tilde{z}(\omega)]^2 - \frac{1}{\sqrt{2}} \frac{Q_0 F V_{P,C}}{k_0} \frac{1}{\operatorname{Im}\tilde{z}(\omega)} + V_{P,C}, \tag{S14}$$

which we simplify to

$$V_P = u\Lambda^2 - \frac{v}{\Lambda} + V_{P,C} \tag{S15}$$

with $u$, $v$, and $V_{P,C}$ as fitting parameters. Here, we make use of the relations $\Lambda = |Y_{pumped}/Y_{unpumped}|$ and $Y \propto \operatorname{Im}[\tilde{z}(\omega)]$ to write

$$\operatorname{Im}[\tilde{z}(\omega)] = \Lambda \cdot \operatorname{Im}[\tilde{z}(\omega)]_{unpumed} \tag{S16}$$

so that $u = \frac{\pi \eta Q_0 f_0 V_{P,C}}{k_0} \operatorname{Im}[\tilde{z}(\omega)]^2_{unpumped}$ and $v = \frac{Q_0 F V_{P,C}}{\sqrt{2} k_0 \operatorname{Im}[\tilde{z}(\omega)]_{unpumped}}$. Please note that the value of $V_{P,C}$ is independent of any renormalization of the motion amplitude.

**D) Additional measurements of self-oscillations.**

We present self-oscillation measurements at $300$ K in Fig. S1. The device is the same as that in the main text, but measured at a time when mechanical and electrical characteristics were different: namely, the conductance is larger by $20$ % and $df_0/dV_g$ of the first mechanical mode is higher ($7$ MHz/V). In addition, the measurements are performed at a different gate voltage ($-1.9$ V). The quality factor obtained from the self-oscillation threshold is $\sim 230$. This is much larger then the quality factor determined from the lineshape of the driven resonance, which is $10$-$15$ (Fig. S1b).

Figure S2 shows measurements from a second carbon nanotube resonator. The quality factor obtained from the self-oscillation threshold is $\sim 1000$. This is again much larger than the quality factor determined from the lineshape of the driven resonance, which is about $100\text{-}220$ (Fig. S2c).

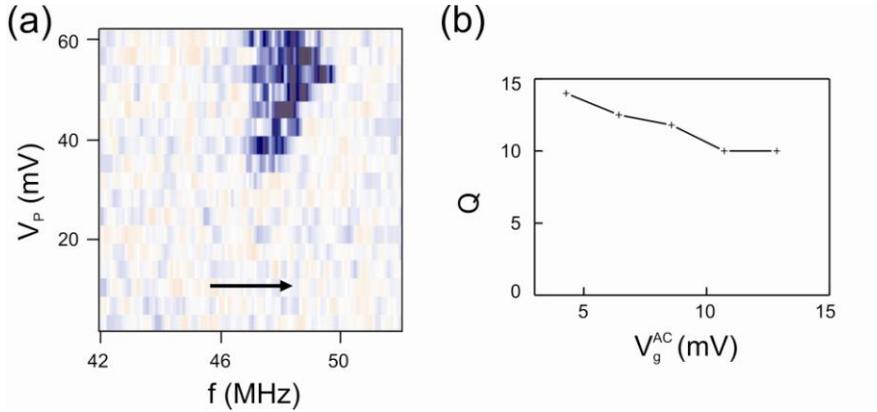

**Figure S1: (a)** Self-oscillations at $300$ K and with $V_g = -1.9$ V. Here, $f_0 = 48$ MHz, $df_0/dV_g = 7$ MHz/V, and $V_{P,C} \sim 30$ mV. The corresponding quality factor is $\sim 230$. **(b)** Quality factor as a function of the driving voltage $V_g^{AC}$ in the absence of parametric pumping, obtained by fitting the resonance lineshape with the predictions of a damped harmonic oscillator.

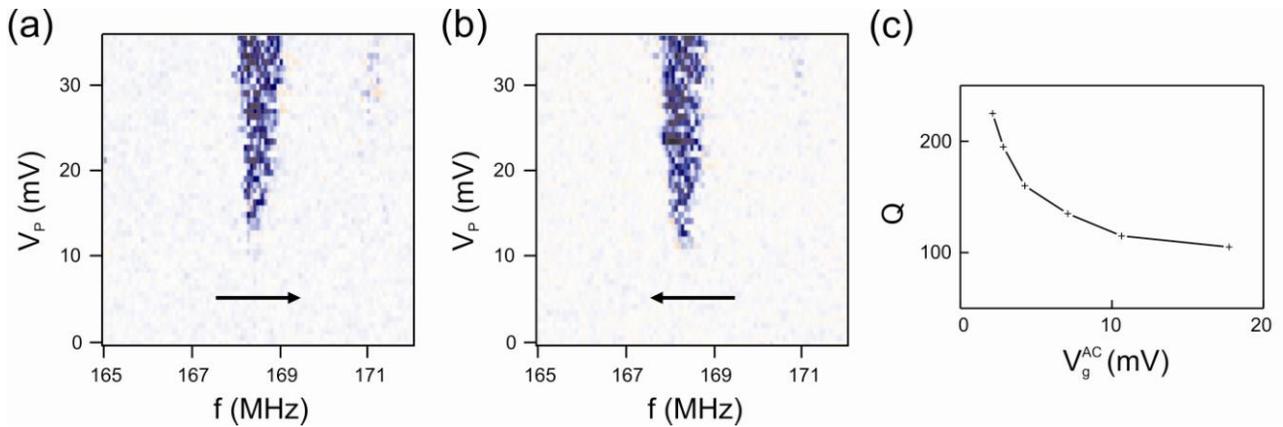

**Figure S2:** Data from a second nanotube device at $T = 60$ K and $V_g = 1.8$ V **(a) and (b)** Self-oscillation with increasing and decreasing frequency sweeps, respectively. Self-oscillations are detected above $V_{P,C} = 10$ mV in a tongue-shaped region, which corresponds to a quality factor of $\sim 1000$ ($f_0 \sim 168$ MHz and $df_0/dV_g = 14.2$ MHz/V). In contrast to the data shown in Fig. 3 of the main text, no hysteresis is observed. **(c)** Quality factor as a function of the driving voltage $V_g^{AC}$ in the absence of parametric pumping, obtained by fitting the resonance lineshape with the predictions of a damped harmonic oscillator.